\pdfoutput=1

\documentclass{article}
\usepackage[T1]{fontenc}
\usepackage[utf8]{inputenc}
\usepackage{ismir}
\usepackage{amsmath,amssymb,cite,url}
\usepackage{graphicx}
\usepackage{color}
\usepackage{booktabs}
\usepackage{multirow}
\usepackage{tikz}
\usetikzlibrary{arrows.meta,positioning,calc,fit}

\title{Learning Jazz Pianist Style with \\ Cross-Attention Conditioning}

\threeauthors
  {Drew Edwards}
  {Centre for Digital Music,\\Queen Mary University of London\\\texttt{a.c.edwards@qmul.ac.uk}}
  {Akira Maezawa}
  {Yamaha Corporation\\Hamamatsu, Japan}
  {Simon Dixon}
  {Centre for Digital Music,\\Queen Mary University of London}

\def\authorname{D. Edwards, A. Maezawa, and S. Dixon}

\begin{document}

\maketitle

\begin{abstract}
Jazz pianists develop distinctive traits that experienced listeners
can often identify within seconds, yet the features underlying this
recognition resist formal description. We study jazz pianist style
through the lens of a pretrained symbolic music transformer, showing
that its learned representations already encode pianist identity
well enough for highly accurate classification across two
benchmarks. We then augment the transformer with cross-attention
over learned pianist identity embeddings, enabling it to generate
music conditioned on a specific artist's style. Two evaluation
protocols confirm that the generator captures meaningful stylistic
structure: a sliding-window classifier consistently attributes
conditioned continuations to the correct artist, far above
unconditioned baselines; and a classifier trained \emph{entirely}
on synthetic generations identifies real pianists across 12
classes with
87\% chunk-level and 95\% song-level accuracy. Finally, we repurpose the classifier to
locate the most characteristic moments within a performance,
surfacing the specific musical gestures that distinguish each
pianist's voice.
\end{abstract}

\section{Introduction}\label{sec:intro}

An experienced jazz listener can identify Art Tatum or Bill Evans
within a few bars, yet the features that make their playing so
recognizable resist precise
description~\cite{berliner1994thinking}. \emph{DownBeat}
magazine's Blindfold Test---in which jazz musicians identify
performers from recordings---has demonstrated this perceptual
ability for over seven decades. Jazz pianists also directly imitate the greats in order to
understand and incorporate their playing styles. Dick Hyman's
\emph{In the Styles Of\ldots\ The Great Jazz
Pianists}~\cite{hyman1994styles} is a
collection of original etudes, each written to embody a different
pianist's voice. Hyman describes Tatum's ``perfect legato'' and
``rapid runs in both hands,'' Garner's ``solid touch'' and
``strumming, guitar-like left hand,'' and Peterson's ``tremolos
and glissandi'' and ``single-note lines interspersed with thirds.''
His work poses a natural question for music generation research:
can a model learn to internalize and reproduce these stylistic
signatures?

Symbolic music generation systems can now produce convincing piano
music~\cite{huang2019music,bradshaw2025scaling}, but conditioning
on \emph{individual performer identity}---rather than genre, era,
or mood---remains largely unexplored. We study pianist style at
three levels: embedding analysis reveals geometric structure in how
pianists cluster (Section~\ref{sec:data}); classification confirms
that style can be automatically distinguished
(Section~\ref{sec:approach}); and conditional generation tests
whether it can be synthesized
(Sections~\ref{sec:persistence}--\ref{sec:characteristic}). A key
challenge is that standard likelihood metrics are nearly blind to
conditioning quality---models with and without artist conditioning
produce near-identical perplexities
(Table~\ref{tab:perplexity}). This motivates classifier-based
evaluation that directly probes whether generated music carries
artist identity.

We make the following contributions. First, we show that an
Aria-based~\cite{bradshaw2025scaling} pianist classifier exceeds
the state of the art on the Deep Pianist Identification
benchmark~\cite{cheston2025deeppianist}
(Section~\ref{sec:approach}). Second, we introduce a gated
cross-attention conditional generator and evaluate it with two
style-fidelity tests: a sliding-window agreement protocol showing
a sustained $\sim$33-point advantage over baselines
(Section~\ref{sec:persistence}), and a synthetic-transfer
test~\cite{ravuri2019cas} in which a 12-way classifier trained
\emph{only} on generated music reaches 95\% song-level accuracy on
real performances (Section~\ref{sec:synthetic}). Third, we
repurpose the classifier to locate the most stylistically
distinctive moments within a performance
(Section~\ref{sec:characteristic}). Audio demos and more can be found at the project webpage. \footnote{\tiny\url{https://almostimplemented.github.io/jazz-pianist-style}}

\section{Related Work}\label{sec:related}

\textbf{Symbolic music generation.}
The Music Transformer~\cite{huang2019music} introduced relative
self-attention, where attention weights depend on the distance
between musical events rather than their absolute position,
enabling the model to capture repetition and phrase structure.
MuseNet~\cite{payne2019musenet} demonstrated composer-level
conditioning by prepending genre, composer, and instrumentation
tokens to the input sequence of a large-scale transformer. Oore et
al.~\cite{oore2020time} pioneered expressive piano generation with
an event-based MIDI representation encoding note onsets, offsets,
velocities, and time shifts as discrete tokens---the same
representational family that Aria builds on. More recently,
MuseMorphose~\cite{wu2023musemorphose} applied a Transformer VAE
to fine-grained piano style transfer at the bar level, and Zhang
et al.~\cite{zhang2024composer} conditioned symbolic generation on
composer identity via VQ-diffusion.
Aria~\cite{bradshaw2025scaling} is a self-supervised piano
transformer whose contrastive variant produces the per-track embeddings
that we use for artist selection and classification. While
score-to-performance rendering systems have conditioned on
individual pianist identity to predict expressive
parameters~\cite{tang2023reconstructing,zhang2023horowitz},
conditioning a generative model on performer identity to produce
the musical content itself remains underexplored. Jazz makes this
particularly compelling: unlike classical piano, where different
performers interpret the same score and style resides primarily in
expressive nuance, jazz pianists are simultaneously composers and
performers. Their harmonic choices, melodic vocabulary, and
rhythmic patterns are as much a part of their identity as their
touch or phrasing. Modeling jazz pianist style therefore requires
generating the musical content itself, not only rendering a given
score expressively.

\textbf{Performer identification.}
Cheston et al.~\cite{cheston2024rhythmic} showed that rhythmic
features extracted from source-separated audio---including swing
ratio, tempo variability, and micro-timing deviations---predict
jazz pianist identity well above chance;
their subsequent Deep Pianist Identification
benchmark~\cite{cheston2025deeppianist} achieved 94.4\% track-level
accuracy with a ResNet-50 over piano-roll features. The PiJAMA
dataset~\cite{edwards2023pijama} provides aligned audio and MIDI
for 30 jazz pianists. We build on both resources.

\looseness=-1\relax\textbf{Cross-attention conditioning.}
Flamingo~\cite{alayrac2022flamingo} introduced gated
cross-attention layers for conditioning a frozen language model on
a new modality; IP-Adapter~\cite{ye2023ipadapter} applied the
pattern to image generation. We adopt the same gated-residual
design but with a simpler conditioning source: a small learned
per-artist embedding matrix injected into a subset of the
model's layers via cross-attention.

\looseness=-1\relax\textbf{Evaluating conditional generation.}
Teacher-forced perplexity is standard but, as we show, nearly
insensitive to conditioning quality. Distributional metrics such
as Fr\'{e}chet Audio Distance~\cite{kilgour2019fad} measure
aggregate similarity but do not test whether individual samples
carry the intended conditioning. Classifier-based evaluation has
been explored in vision~\cite{lesort2019training}; the
Classification Accuracy Score (CAS)~\cite{ravuri2019cas}---training
a classifier on generated data and testing on real data---directly
inspires our synthetic transfer experiment
(Section~\ref{sec:synthetic}). High accuracy implies the generator
has encoded class-discriminative structure that transfers to real
examples.

\section{Data and Artist Selection}\label{sec:data}

\begin{figure}[t]
  \centering
  \includegraphics[alt={t-SNE of Aria-CL embeddings for 30 PiJAMA pianists with selected 12 highlighted.},width=0.82\linewidth]{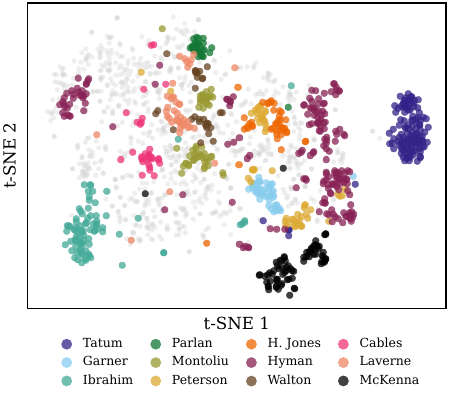}
  \caption{t-SNE~\cite{vandermaaten2008tsne} projection of Aria-CL
    track embeddings for all PiJAMA-30 pianists. The 12 selected
    pianists (colored) occupy distinct clusters; remaining artists
    are shown in light gray.}
  \label{fig:tsne}
  \vspace{-5pt}
\end{figure}

We draw from PiJAMA-30~\cite{edwards2023pijama}, a dataset of solo
jazz piano performances spanning 30 artists with aligned audio and
automatically transcribed MIDI. For this study, we select a subset
of the 12 most distinctive pianists, which we call PiJAMA-12,
focusing on artists whose styles are sufficiently differentiated
for both a classifier and a conditional generator to exploit.
To identify these artists, we compute per-track embeddings using
the contrastive variant of Aria~\cite{bradshaw2025scaling} and
rank all 30 PiJAMA artists by $k$NN leave-one-out accuracy and
silhouette score~\cite{rousseeuw1987silhouettes}, selecting the
top~12. The selected pianists---Art Tatum, Erroll Garner, Abdullah
Ibrahim, Horace Parlan, Tete Montoliu, Oscar Peterson, Hank Jones,
Dick Hyman, Cedar Walton, George Cables, Andy Laverne, and Dave
McKenna---occupy well-separated regions in embedding space
(Figure~\ref{fig:tsne}), spanning a range of eras and stylistic
approaches. This selection prioritizes stylistic separability over
coverage; scaling to less separable artists is a natural next step.

After removing 22 PiJAMA-30 recordings that contain ensemble
playing rather than solo piano, we tokenize all MIDI using Aria's
AbsTokenizer~\cite{bradshaw2025scaling}, with three tokens (pitch-velocity, onset, duration) per note, producing sequences
of 4096 tokens (${\sim}$1\,200 notes) for generation and 1024
tokens (${\sim}$300 notes) for classification. All splits are at the song level
to prevent leakage between train, validation, and test sets. The
resulting PiJAMA-12 subset contains 779 songs ($\sim$57 hours)
across the 12 artists, with per-artist counts ranging from 28
(Horace Parlan) to 187 (Dick Hyman), split into 1\,383 training,
175 validation, and 177 test sequences (at 4096 tokens; the
classifier uses these same songs re-chunked into 1024-token
windows).

\section{Approach}\label{sec:approach}

\subsection{Pianist Classification}\label{sec:classifier}

We train a pianist classifier by fine-tuning the full Aria medium
backbone (659M parameters; 16 layers, $d{=}1536$) with a linear
classification head over the hidden state at the end-of-sequence
token. Training uses class-balanced cross-entropy with pitch,
tempo, and velocity augmentation~\cite{bradshaw2025scaling}
(learning rate $10^{-5}$, batch size 16, 15 epochs).

To validate the feature backbone, we first evaluate on the Deep
Pianist Identification benchmark~\cite{cheston2025deeppianist}
using their published splits and 20-class setup. Our classifier
reaches 96.9\% track-level accuracy, surpassing their ResNet-50
baseline of 94.4\% (Table~\ref{tab:dpi}). We then train the
classifier that serves as our evaluation tool throughout
Sections~\ref{sec:persistence} and~\ref{sec:synthetic} on the
PiJAMA-12 subset, achieving 95.8\% chunk-level and 98.8\%
track-level accuracy on the held-out test split. An identical
recipe applied to all 30 PiJAMA artists reaches 90.2\% and 99.4\%,
respectively (Table~\ref{tab:dpi}).

\begin{table}[t]
  \centering
  \small
  \begin{tabular}{lrrr}
    \toprule
    \textbf{Method / Dataset} & \textbf{Chunk} & \textbf{Clip} & \textbf{Track} \\
    \midrule
    ResNet-50~\cite{cheston2025deeppianist} on DPI-20 & --- & 80.5 & 94.4 \\
    Ours (Aria) on DPI-20     & 84.1 & 90.4 & \textbf{96.9} \\
    \midrule
    Ours on PiJAMA-12         & 95.8 & ---  & 98.8$^\dagger$ \\
    Ours on PiJAMA-30         & 90.2 & ---  & 99.4$^\dagger$ \\
    \bottomrule
  \end{tabular}
  \caption{Pianist identification accuracy (\%). A \emph{chunk} is
    a 1024-token classification window; a \emph{clip} is a
    $\sim$30\,s audio segment following the protocol
    of~\cite{cheston2025deeppianist}. Top: Deep Pianist benchmark
    (20 classes) using published splits; clip and track columns
    aggregate windows by mean logit.
    Bottom: PiJAMA classifiers (12 and 30 classes,
    identical recipe) on held-out song-level test splits (80 and
    156 songs); the 12-class model is used for all style-fidelity
    evaluations.
    $^\dagger$Majority vote, ties broken by mean logit; each
    classifier misses one short track (of 80 and 156).}
  \label{tab:dpi}
  \vspace{-6pt}
\end{table}

\subsection{Cross-Attention Conditional Generation}\label{sec:crossattn}

\begin{figure}[t]
  \centering
  \includegraphics[alt={Gated cross-attention block diagram.},width=0.65\linewidth]{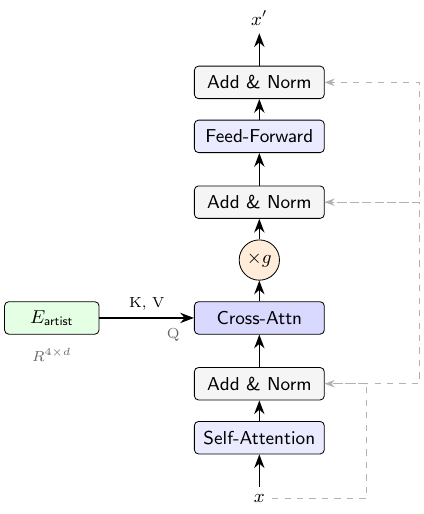}
  \caption{Gated cross-attention block added to the last 8 of 16
    transformer layers. Queries come from the music sequence; keys
    and values from learned artist embeddings.}
  \label{fig:architecture}
\end{figure}

Starting from the pretrained Aria medium model, we insert a gated
cross-attention layer after the self-attention block in the
last 8 of 16 transformer layers (Figure \ref{fig:architecture}). Queries come from the music
sequence; keys and values come from a learned artist embedding
$E_{\text{artist}} \in \mathbb{R}^{4 \times d}$:
\begin{equation}
  x' = x + g \cdot \mathrm{CrossAttn}\!\left(x,\, E_{\text{artist}}\right)
  \label{eq:cross_attn}
\end{equation}
where $g$ is a scalar gate initialized to $0.1$, following the
Flamingo design~\cite{alayrac2022flamingo}. This ensures the
adapter starts as a near-identity residual. Unlike prompt-based
conditioning, which loses influence as generation advances beyond
the initial context~\cite{vonrutte2023figaro}, cross-attention
provides a persistent channel that does not decay with sequence
length. The 8 cross-attention
layers and 12 artist embeddings add 78M parameters to the 659M
base model (12\% overhead). Element-wise dropout ($p{=}0.3$) is
applied to the projected artist context during training as
regularization.

We fine-tune the full model plus adapter for 15 epochs on
PiJAMA-12 training sequences of 4096 tokens using AdamW (weight
decay $0.02$, 0.5-epoch linear warmup, effective batch size 32,
peak learning rate $5{\times}10^{-6}$, fp16) on a single NVIDIA
RTX~6000 GPU in approximately 18 hours. As an ablation control, we
also train a baseline with identical hyperparameters but \emph{no}
cross-attention layers.

\looseness=-1 Table~\ref{tab:perplexity} reports token-weighted test-set
perplexity. Fine-tuning closes nearly all of the
pretrained-to-in-domain gap; cross-attention adds only a marginal
further improvement. This is consistent with the well-known
difficulty of training global conditioning signals in powerful
autoregressive decoders~\cite{bowman2016generating}: under teacher
forcing, the ground-truth prefix provides sufficient local context
for accurate next-token prediction, leaving little room for a
global signal such as pianist identity to improve the loss. The
conditioning signal's effect surfaces only during free generation,
when the model must sustain stylistic coherence from its own
outputs. This motivates the classifier-based evaluations in the
following sections.

\begin{table}[t]
  \centering
  \small
\addtolength{\tabcolsep}{-3pt}
\begin{tabular}{lrrr}
    \toprule
    \textbf{Model} & \textbf{Test NLL} & \textbf{Perplexity} & \textbf{Agreement}$^\dagger$ \\
    \midrule
    Pretrained Aria            & 2.43 & 11.41 & 25\% \\
    Fine-tuned (no CA)         & 1.94 & 6.96  & 37\% \\
    Cross-attention (ours)     & \textbf{1.92} & \textbf{6.82} & \textbf{70\%} \\
    \bottomrule
  \end{tabular}
  \caption{Token-weighted cross-entropy on the PiJAMA-12 test
    split (177 sequences, 562\,125 tokens) and mean classifier
    agreement during free generation. Despite near-identical
    perplexities, style fidelity differs dramatically.
    $^\dagger$Mean sliding-window agreement at $P{=}256$
    (Figure~\ref{fig:drift}).}
  \label{tab:perplexity}
  \vspace{-6pt}
\end{table}

\section{Style Persistence Evaluation}\label{sec:persistence}

Since teacher-forced perplexity barely distinguishes our models, we
introduce a classifier-based protocol that operates on the model's
own autoregressive output. For each validation sample with true
artist $A$, we extract a prompt of length $P$ tokens and generate a
continuation of up to 4096 tokens. A 1024-token classification
window is slid across the continuation at stride 128, and we
record the fraction of windows where the classifier's top-1
prediction matches $A$, as a function of window start position.
We call this fraction the \emph{agreement} with $A$
(Table~\ref{tab:perplexity} reports its positional mean).

We evaluate four generation modes: (i)~\emph{conditioned}
(cross-attention with true artist context), (ii)~\emph{ablation}
(same architecture, conditioning context set to zero),
(iii)~\emph{fine-tuned baseline} (no cross-attention layers), and
(iv)~\emph{pretrained Aria} (no fine-tuning). All modes use
temperature $0.95$ with top-$k{=}50$ and top-$p{=}0.95$.

\begin{figure*}[t]
  \centering
  \includegraphics[alt={Sliding-window classifier agreement vs position for four modes at three prompt lengths.},width=0.74\linewidth]{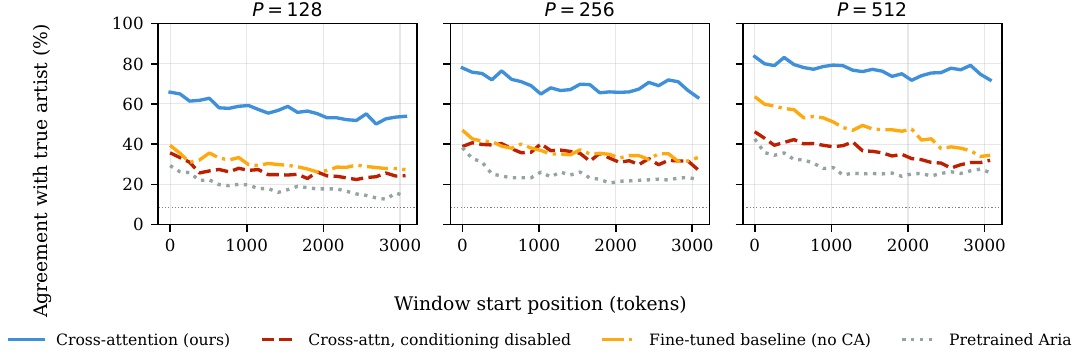}
  \caption{Classifier agreement with the true artist as a function
    of window-start position along a 4096-token continuation.
    Three panels show prompt lengths $P \in \{128, 256, 512\}$. The
    conditioned model (blue) sustains a $\sim$33-point advantage
    over the fine-tuned baseline across the full generation. The
    ablation (dashed red) tracks the fine-tuned baseline, isolating
    the effect of the conditioning signal from the architecture.}
  \label{fig:drift}
  \vspace{-9pt}
\end{figure*}

Figure~\ref{fig:drift} shows the results for $P \in
\{128, 256, 512\}$ evaluated over the 175 validation sequences.
The conditioned model maintains a large and stable advantage across
the full 4096-token generation: at $P{=}256$, it averages 70\%
agreement with the true artist, versus 37\% for the fine-tuned
baseline and 25\% for pretrained Aria. The
ablation run---which uses the cross-attention architecture but
receives no conditioning context---tracks the fine-tuned baseline
almost exactly, confirming that the gap is due to the conditioning
signal, not the additional parameters. Notably, the conditioned
curve decays at a similar rate to the baselines rather than
remaining flat, suggesting the prompt provides complementary
stylistic information that the conditioning signal does not fully
replace. The conditioning signal's influence is bounded by the
learned gate $g$ (Equation~\ref{eq:cross_attn}), which balances
style fidelity against musical coherence: a gate too large would
impose the artist embedding at the expense of continuity.
The result is a stable \emph{offset} in agreement rather than a
different decay rate.

\begin{figure}[t]
  \centering
  \includegraphics[alt={Per-artist conditioned classifier agreement at P=256.},width=0.83\linewidth]{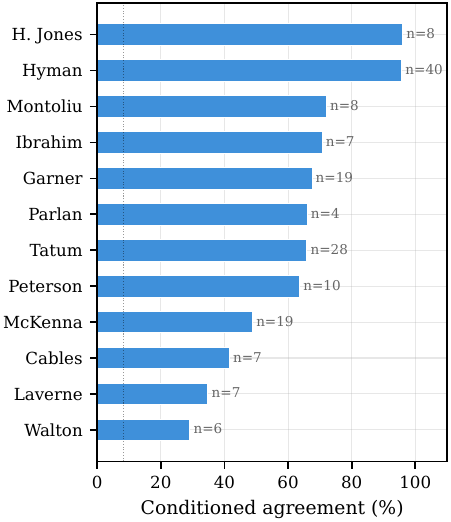}
  \caption{Per-artist conditioned agreement at $P{=}256$
    (Section~\ref{sec:persistence}). Sample counts shown at right.
    Agreement varies from 96\% (Hank Jones) to 29\% (Cedar Walton);
    the dotted line marks chance (8.3\%).}
  \label{fig:per_artist_agreement}
\end{figure}

Figure~\ref{fig:per_artist_agreement} shows the per-artist
breakdown at $P{=}256$. Agreement varies widely, from 96\% (Hank
Jones, Dick Hyman) to 29\% (Cedar Walton). This variation does not
simply reflect training data volume: Hank Jones achieves 96\% from
just 69 training sequences, while Dave McKenna reaches only 49\%
from nearly twice as many. Notably, Andy Laverne achieves high
classification accuracy on real data
(Figure~\ref{fig:per_artist_acc}) but only 35\% conditioned
agreement, indicating that a recognizable style does not guarantee
the generator can reproduce it.

We also test a \emph{mismatch} condition: prompting from artist
$A$ but conditioning on artist $B \neq A$. In individual
generations, the classifier's attribution shifts cleanly from the
prompt artist to the conditioning artist partway through the
continuation. In aggregate across 148 mismatch pairs at
$P{=}512$, the prompt artist's agreement decays from 40\% to
27\% while the conditioning artist's rises from 15\% to 24\%,
with the two converging near position $\sim$1500---confirming that
the conditioning signal gradually overrides the prompt's
stylistic influence.

\section{Synthetic Transfer Evaluation}\label{sec:synthetic}

\looseness=-1 The sliding-window test measures whether a \emph{pretrained}
classifier recognizes artist identity in generated output. A more
demanding test removes any dependence on the pretrained classifier:
we train a \emph{new} classifier exclusively on
generated MIDI, and evaluate it on real performances it has never
seen.

\textbf{Protocol.} We take each PiJAMA-12 training sequence, use
the first 256 tokens as a prompt, and generate a 4096-token
continuation conditioned on the true artist. The prompt tokens are
discarded; only the continuation is retained. These are re-chunked
into 4\,782 sequences of 1024 tokens. An Aria-based
classifier (same setup as
Section~\ref{sec:classifier}) is trained exclusively on this
synthetic data and evaluated on the real PiJAMA-12 test set.

\begin{figure}[t]
  \centering
  \includegraphics[alt={Per-artist chunk accuracy for real and synthetic classifiers.},width=0.82\linewidth]{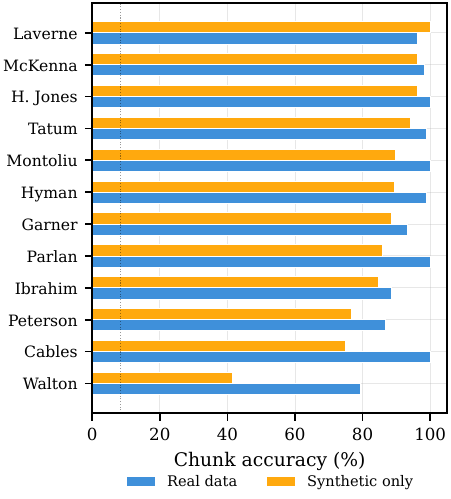}
  \caption{Per-pianist chunk-level test accuracy on PiJAMA-12.
    Blue: real-data classifier (Section~\ref{sec:classifier}).
    Orange: synthetic-only classifier (Section \ref{sec:synthetic}). Artists are
    ordered by synthetic accuracy.}
  \label{fig:per_artist_acc}
  \vspace{-5pt}
\end{figure}

\textbf{Results.}\looseness=-1\relax\ Table~\ref{tab:synth_transfer} shows that the
synthetic-only classifier reaches 87.1\% chunk-level and 95.0\%
song-level accuracy on real performances---within 9 chunk-level
points of the real-data classifier---demonstrating that the
generated MIDI encodes stylistic structure that transfers to real
recordings.
To isolate the classifier's dependence on Aria pretraining, we repeat
the experiment with a classifier that has no pretraining at all:
a ResNet-50 over piano-roll clips, trained from random
initialization with the architecture and recipe
of~\cite{cheston2025deeppianist} and model selection on held-out
generations. During training it never sees any MIDI data other
than the output of our generative model---yet it identifies
91.3\% of real test tracks (70.2\% of 30\,s clips;
Table~\ref{tab:synth_transfer}): recognizing the artist in the
generated data requires no Aria-pretrained representations.
Trained instead on the real training split, the same network
reaches 97.5\% of tracks (93.1\% of clips); the synthetic-only
ResNet is thus within 7 track-level points of its real-trained
counterpart.
Figure~\ref{fig:per_artist_acc} breaks down chunk-level
accuracy by artist: the gap between real and synthetic classifiers
is largest for Cedar Walton and George Cables, who also have some of the lowest conditioned agreement
(Figure~\ref{fig:per_artist_agreement}).

\textbf{Memorization and diversity.} Three potential confounds
deserve attention. First, since the generator was trained on the
same data used as prompts, it might simply replay memorized
sequences. Pitch and interval $n$-gram overlap between generated
continuations and training data drops below 5\% at $n{=}6$ and to
1\% or below at $n{=}10$, and a stricter windowed piano-roll
similarity~\cite{yin2022originality,ji2025ssimuse} agrees:
generations match their closest training track \emph{less}
closely than held-out real performances do (mean max-Jaccard 0.20
vs.\ 0.32 over 8\,s windows, excluding each continuation's own
source recording), and no generation is anomalously close to any
single unrelated training or test track. Second, the generator
might collapse to a single stereotyped output per artist.
Pairwise $4$-gram Jaccard similarity between same-artist
generations is $1.5$--$3.5\times$ higher than between-artist
generations---matching the ratio observed in real data---with
absolute values far below 1.0. Third, Aria's pretraining corpus
may share source recordings with PiJAMA. Behaviorally, pretrained
Aria shows no replay signature on our evaluation prompts---its
unconditioned continuations collapse toward the corpus prior, not
the true artist---and the transfer result survives with a
classifier that has no pretraining at all; the two corpora were
also transcribed by different systems, so even shared recordings
would not yield identical token sequences.

\begin{table}[t]
  \centering
  \small
  \begin{tabular}{lrr}
    \toprule
    \textbf{Training data} & \textbf{Chunk} & \textbf{Song}$^\dagger$ \\
    \midrule
    Real PiJAMA-12         & 95.8 & 98.8 \\
    \ + from scratch (ResNet-50) & 93.1$^\ddagger$ & 97.5 \\
    Synthetic only (ours)  & 87.1 & 95.0 \\
    \ + from scratch (ResNet-50) & 70.2$^\ddagger$ & 91.3 \\
    \bottomrule
  \end{tabular}
  \caption{Pianist identification on the \emph{real} PiJAMA-12 test
    set (80 songs, 591 chunks). Synthetic-trained classifiers see
    only generated continuations; within each architecture, the
    real- and synthetic-trained classifiers share one training
    recipe.
    $^\dagger$Majority vote over chunks within a song; exact vote
    ties are broken by mean logit.
    $^\ddagger$30\,s piano-roll clips rather than token chunks.}
  \label{tab:synth_transfer}
  \vspace{-6pt}
\end{table}

\begin{figure*}[t]
  \centering
  \includegraphics[alt={Characteristic region detection for Art Tatum performing Sophisticated Lady. Top: z-score curve over the full performance. Bottom: piano rolls of the most and least characteristic regions colored by z-score.},width=0.95\linewidth]{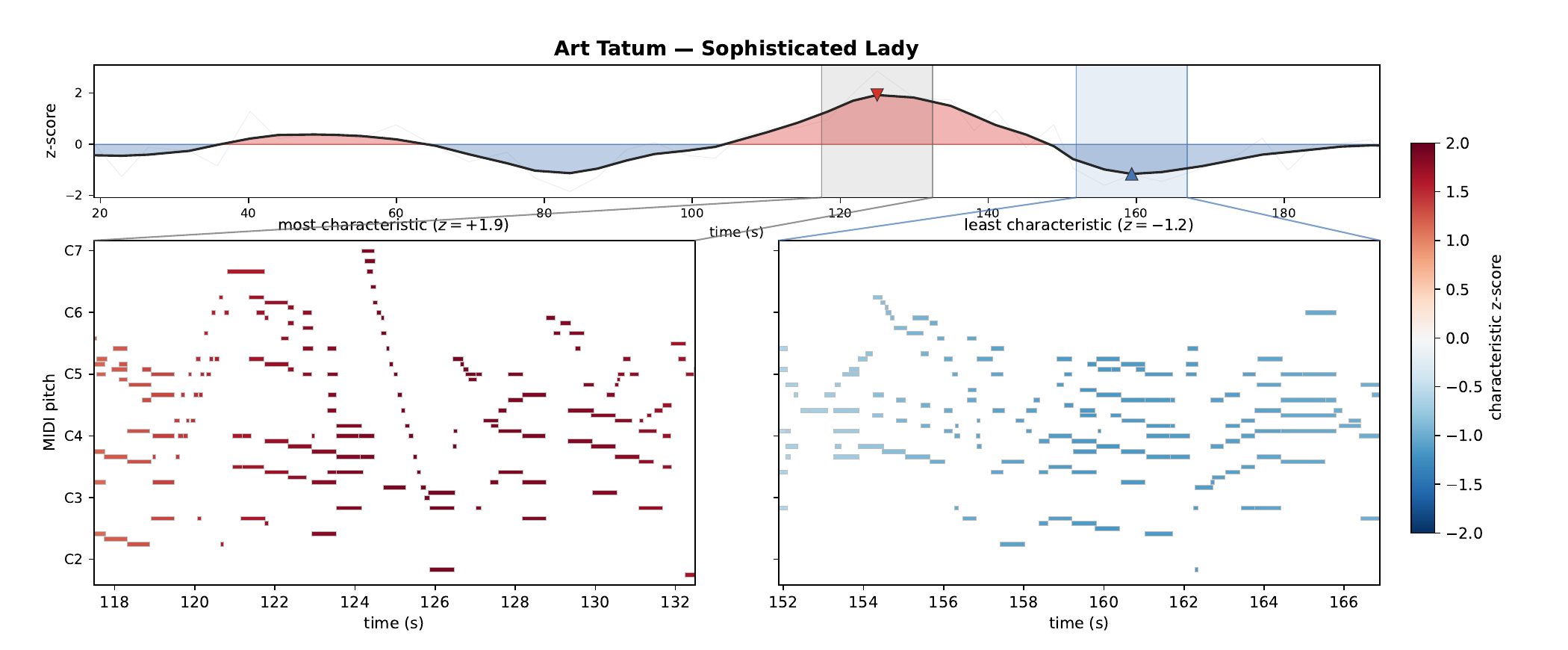}
  \caption{Characteristic region detection for Art Tatum,
    ``Sophisticated Lady.'' Top: within-track z-score of the
    classifier's logit margin; red fill marks confidence above the
    track average, blue below. Bottom: piano rolls of the most
    (left, $z{=}1.9$) and least (right, $z{=}{-}1.2$)
    characteristic 15-second regions, notes colored by z-score.
    The peak's dense runs span over five octaves, consistent with
    Tatum's virtuosic runs; the trough shows
    comparable note density (9.7 vs.\ 10.6 notes/s) over a
    narrower range (4.4 vs.\ 5.2 octaves), suggesting the
    classifier's confidence tracks specific gestures, not overall
    activity.}
  \label{fig:characteristic}
  \vspace{-9pt}
\end{figure*}

\section{Characteristic Region Detection}\label{sec:characteristic}

The evaluations in Sections~\ref{sec:persistence}
and~\ref{sec:synthetic} measure style fidelity in aggregate---across
entire continuations or entire test sets. A complementary question
is whether we can identify \emph{where} in a performance the
features that define a pianist's style are most concentrated.

The classifier built in Section~\ref{sec:classifier} assigns a
single label to each chunk, but a performance is not uniformly
characteristic: some passages may be more distinctively ``Art
Tatum'' than others. By sliding the classifier across a held-out
performance, we can attempt to locate these moments. A na\"{\i}ve
approach would track the classifier's raw confidence, but at
95.8\% chunk-level accuracy the classifier is already near-certain
on almost every window---raw logit margins hover around
$15 \pm 0.5$ for a typical track, leaving almost no ranking
signal. We therefore ask a relative question: where is the
classifier \emph{even more confident than usual} for this
particular performance?

For each 1024-token window $w$ at stride 128, we compute the
\emph{logit margin} between the true artist and the strongest
alternative:
\begin{equation}
  m(w) \;=\; \ell_{\text{true}}(w) \;-\; \max_{j \,\neq\, \text{true}}\, \ell_j(w)
  \label{eq:margin}
\end{equation}
We then normalize within each track:
\begin{equation}
  z(w) \;=\; \frac{m(w) - \mu_T}{\sigma_T}
  \label{eq:zscore}
\end{equation}
where $\mu_T$ and $\sigma_T$ are the mean and standard deviation
of $m$ across all windows in track~$T$. Intuitively, $z(w)$
measures how much more confidently the classifier recognizes the
true pianist in window~$w$ compared to its average confidence
across the whole performance. Gaussian smoothing
($\sigma{=}1.5$~windows) produces a continuous curve.

Figure~\ref{fig:characteristic} illustrates the result for Art
Tatum's ``Sophisticated Lady.'' The z-score curve (top) shows that
most of the performance sits near zero---not because the classifier
is uncertain, but because its high confidence is uniform. A peak
near $t{=}125$\,s stands out: the piano roll (bottom) shows this
region contains dense note activity spanning over five octaves,
consistent with the ``rapid runs in both hands'' that Hyman
highlights in his Tatum etude (Section~\ref{sec:intro}).

Across all 80 test tracks (12 artists), the median peak z-score
is 0.90 and the maximum is 1.99 (Art Tatum). The method reveals
distinct patterns across artists. Artists like Art Tatum produce
clear peaks, suggesting their style concentrates in specific
gestures that the classifier finds especially discriminative. By
contrast, Erroll Garner produces no comparable peaks across any of
his four test tracks (maximum $z{=}0.59$), despite the classifier
identifying him with equally high confidence on every window. For
Garner, the features the classifier relies on appear to be
distributed uniformly throughout a performance rather than
concentrated in discrete moments.

\section{Discussion and Conclusion}\label{sec:discussion}

A recurring theme is the disconnect between token-level and
sequence-level evaluation. Under teacher forcing, the ground-truth
prefix provides rich local context that is highly predictive of
the next token regardless of who is playing, leaving little room
for a global signal like artist identity to improve the loss
(Table~\ref{tab:perplexity}). During free generation, however,
small per-token biases accumulate: without conditioning, stylistic
coherence degrades; with it, the artist embedding provides a
persistent anchor. Pianist style is thus a \emph{sequence-level}
property, and our style-fidelity tests
(Table~\ref{tab:perplexity}, rightmost column) confirm that the
conditioning signal carries information invisible to perplexity.

Some pianists consistently emerge as ``easy'' and ``hard''
across evaluations. Per Figure~\ref{fig:per_artist_agreement},
Hank Jones and Dick Hyman
achieve ${>}$95\% conditioned agreement while Cedar Walton and
Andy Laverne fall below 35\%. Classification on real performances
shows a similar pattern---Horace Parlan, who has the fewest
training samples, achieves 100\% chunk-level accuracy
(Figure~\ref{fig:per_artist_acc}). These results suggest that
\emph{stylistic distinctiveness}, rather than data quantity, is the
primary factor: pianists with a consistent and recognizable voice
are both easier to classify and to imitate.

The concentration of misclassifications on Dick Hyman is
notable: for Art Tatum, 82\% of incorrectly classified windows
are attributed to Hyman; for Erroll Garner, 86\%. This may
reflect classifier bias toward the most-represented training
class, or it may stem from Hyman's mastery of various
styles, making him a natural ``default'' for ambiguous passages.
By contrast, George Cables' errors distribute across multiple
artists, suggesting a genuinely less distinctive generated style.

Several directions remain open. Combining artist conditioning
with musical structure---lead sheets, chord progressions, or
melodic constraints---would enable ``in-the-style-of'' generation
over specified material. Mixing artist embeddings at inference
could enable interpolation between styles, and classifier-free
guidance~\cite{ho2022classifierfree} could strengthen the
conditioning signal. A human listening study---ideally modeled on \emph{DownBeat}'s
Blindfold Test---would complement our classifier-based evaluations
with perceptual judgments and remains the most important next step.
Scaling beyond 12 artists---and beyond solo piano---would test the
generality of the approach.

Taken together, our results show that a gated cross-attention
adapter captures enough of each pianist's style for a downstream
classifier to learn from its output alone, and that the classifier
itself can be repurposed to locate the musical gestures most
characteristic of each artist. The characteristic-region method
is particularly promising as a bridge between machine learning and
musicology: it can surface aspects of style that are difficult to
articulate formally, offering students and scholars a tool for
understanding what makes each pianist's voice distinctive. More
broadly, these findings suggest that conditional generation models
encode richer stylistic structure than standard evaluation reveals,
and that classifier-based fidelity tests offer a practical means
of accessing it.

\bibliography{references}

\section{AI Usage Statement}

An AI coding assistant (Anthropic Claude) was used throughout
this work to help debug training code, generate figures, interface
with the compute cluster, and write data-processing scripts. AI
was also used as a writing assistant for \LaTeX{} formatting,
tabulation, literature review, and editing.

\section{Ethics Statement}

Style-conditioned music generation raises concerns about
unauthorized imitation. AI-generated music imitating living
artists has already appeared on streaming platforms without
consent~\cite{npr2025aifakes}, causing real harm to musicians'
livelihoods and reputations. Our work is intended as a research
tool for studying and understanding pianist style---a
computational analogue of Dick Hyman's pedagogical
etudes~\cite{hyman1994styles}, not a system for producing
convincing forgeries. We encourage future work in this direction to
consider consent, attribution, and the interests of the artists
whose styles are modeled.

\section{Acknowledgements}
The first author is a research student at the UKRI Centre for
Doctoral Training in Artificial Intelligence and Music, supported
jointly by UK Research and Innovation (grant number EP/S022694/1),
Queen Mary University of London, and Yamaha Corporation.

\end{document}